\documentclass[aps,groupedaddress,superscriptaddress,prd, twocolumn]{revtex4}
\usepackage{hyperref}

\usepackage{amsmath,amssymb,bm}
\usepackage{graphicx}
\usepackage{epstopdf}
\usepackage{amsfonts}
\usepackage{amssymb}
\usepackage{amsbsy}
\usepackage{amsmath}
\usepackage{latexsym}
\usepackage{sansmath}
\usepackage{subfigure}
\usepackage{lipsum}
\usepackage{float}
 \usepackage{bm}
\usepackage{color}
\usepackage{comment}
\usepackage{csquotes}		% for nice looking quotation marks.
\usepackage{physics}
 
\usepackage{accents}
 
\usepackage[plain]{algorithm}
\usepackage{algpseudocode}

\usepackage{xcolor}
\usepackage[normalem]{ulem}

\def\be{\begin{equation}}
\def\ee{\end{equation}}
\def\bea {\begin{eqnarray}}
\def\eea {\end{eqnarray}}
\def\nn {\nonumber}

\def\nn{\nonumber}
\begin{document}
 
%\title{Effective description of the universe}
\title{Quantum cosmology as a lattice in a box}

\author{Mustafa Saeed} \email{mustafa.saeed@centre.edu, msaeed@unb.ca}
\affiliation{Centre College,
Danville, KY, USA 40422}
\affiliation{Department of Mathematics and Statistics, University of New Brunswick, Fredericton, NB, Canada E3B 5A3}
\author{Viqar Husain} \email{vhusain@unb.ca} 
\affiliation{Department of Mathematics and Statistics, University of New Brunswick, Fredericton, NB, Canada E3B 5A3}
  
  \begin{abstract}
\vskip 0.2cm
We describe quantization schemes for scalar field cosmology in the metric variables with fundamental discreteness imposed with a lattice. The variables chosen for quantization determine the lattice, and each lattice produces distinct effective equations derived from  semiclassical states. We show that requiring a bounce at the Planck density uniquely selects the volume lattice and  gives the same effective Friedmann equation as that obtained in loop quantum cosmology. We also present conditions for the validity of the effective equations.

\end{abstract}

\maketitle
\setlength{\arrayrulewidth}{0.1mm}
\renewcommand{\arraystretch}{2}
\numberwithin{equation}{section}

\section{Introduction}
\hspace{\parindent}
%Provide a brief history of the field.
Efforts to quantize gravity date back to the 1930s. The canonical approach encountered conceptual and technical issues such as the problem of time, selection of canonical variables, physical Hilbert space, and interpretation of state vectors of the gravitational field \cite{Dew}. It was realized that these problems can be more easily addressed in symmetry reduced models such as Friedmann-Lema\^{i}tre-Robertson-Walker (FLRW) universe with a scalar field. Early work on such models used the Arnowitt-Deser-Misner (ADM) canonical variables \cite{Dew,BlyIsh}.

%This is \replace{old text}{new text}
%\delete{the}
%This is a check
%This is another check
%\ms{This is a check}

Another method for quantizing cosmological models is  derived from loop quantum gravity (LQG). In LQG, general relativity is expressed in terms of connection-triad canonical variables and the quantum theory is constructed from a representation of the so-called holonomy-flux algebra---the holonomy of the connection on loops and the integral of the triad over surfaces \cite{AshLew}. In LQG it is found that the area operator has a discrete spectrum \cite{RovSmo}. A related quantization of cosmological models - called loop quantum cosmology (LQC) - uses a particular set of reduced holonomy-flux variables \cite{AshPawSi, AshSi}; this quantization is inequivalent to the Schr\"{o}dinger quantization \cite{Halvo,Ashtekar:2002sn} that is usually deployed. A notable success of LQC is the resolution of the big bang singularity via a bounce where matter reaches the Planck scale density.  

The bounce also arises in an effective classical theory \cite{Tav,RovEw} where the Hamiltonian constraint opertator is replaced by its expectation value in a suitably defined semiclassical state \cite{HusWin}. This effective constraint is then used to derive quantum corrected equations of motion.  The effective theory is characterized by the fiducial coordinate volume, a discreteness scale proportional to the Planck scale, and a particular choice of reduced holonomy-flux algebra that is argued to descend from discreteness of area in full LQG.  

In this note we quantize ADM variables for cosmology after discretization on a lattice. Our purpose is to compare the results with LQC, specifically  the choice of variables for quantization, the derivation of the effective Friedmann equation, and the role of fiducial volume \cite{RovEw}. We use scale-invariant and dimensionless variables that remove the fiducial volume at the outset, and use an initially undetermined power of the scale factor as the configuration variable. This facilitates identifying the key features necessary to obtain physically acceptable effective equations in the ADM variables. To derive the effective equations we use the semiclassical states defined in \cite{HusWin}, and note  similarities with work where a direct ``polymerization" is applied to the non-scale invariant ADM variables \cite{Montani_2019,Giovannetti_2022}.

In Section II we  summarize classical cosmology in terms of scale-invariant dimensionless variables; in Sec. \ref{QU} we give the general lattice quantization for any function of the scale factor as the configuration variable; in Sec. \ref{EC} we derive the effective equations from semiclassical states, and the conditions for their validity; lastly in Sec.  \ref{SaD} we give a brief discussion of the main result --- that the effective theory on a volume lattice in the ADM variables is equivalent to the LQC effective theory.

\section{Hamiltonian Cosmology}
\label{ClUn}
 
The canonical action of the FLRW universe minimally coupled to a massless scalar field in the ADM canonical variables is
\begin{equation}
\begin{split}
\label{eq:orac}
S=&V_0\int dt \left[\left(\frac{1}{8\pi G}\right) \dot{a}p_{a} + \dot{\phi}p_{\phi} - N\mathcal{H}\right];\\
%\label{eq:HamConst}
  \mathcal{H}=&-\left(\frac{1}{8\pi G}\right)\left(\frac{p_{a}{}^{2}}{24a}\right)+\frac{p_{\phi}{}^{2}}{2a^3}\approx 0.
\end{split}
\end{equation}
$ V_{0}=\int{d^{3}x} $ is the volume of the cubical fiducial cell (the coordinate box) to which the calculations are restricted, $(a,p_{a})$ and $(\phi,p_{\phi})$ are the phase space variables for gravity and scalar field respectively and $N$ is the lapse function, and overdots indicate derivatives with respect to coordinate time $t$ with metric 
\be
ds^2 = -N^2 dt^2 + a^2(t) (dx^2 + dy^2 + dz^2). 
\ee
The Poisson bracket identified from the action 
\be
\label{eq:pb1}
\{a,p_a\} = \frac{8\pi G}{V_0}
\ee
depends on the fiducial volume but the equations of motion do not; these along with the Hamiltonian constraint lead to the Friedmann equation
\be 
\label{eq:sffe}
\left(\frac{\dot a}{a}\right)^2 =\left(\frac{4\pi G}{3}\right)\rho,  
\ee
where $\rho=p_{\phi}^2/2a^{6}$ is scalar energy density.

%The equations of motion and the Hamiltonian constraint lead to the Friedmann equation
%\be 
%\label{eq:sffe}
%\left(\frac{\dot a}{a}\right)^2 =\left(\frac{4\pi G}{3}\right)\rho.   
%\ee
%where $\rho=p_{\phi}^2/2a^{6}$ is matter energy density; the canonical equations of motion do not depend on $V_0$, since the factors of $V_0$ cancel in the action and the Poisson bracket
%\be
%\{a,p_a\} = \frac{8\pi G}{V_0}.
%\ee

In these canonical variables it appears that the quantum theory will depend on the coordinate volume $V_0$ if the usual quantization rule of converting the Poisson bracket \eqref{eq:pb1} into a commutator is followed. Although this would be acceptable for quantum theory on a fixed background where physical box size is an infrared cutoff, here the physical volume is $a^3V_0$, which is a dynamical variable. Furthermore, the metric is invariant under the scale factor and coordinate rescalings 
\be
a\longrightarrow a/\lambda,\quad x\rightarrow \lambda x
\ee
 and the reduced action  is invariant provided 
 \be
 p_a \rightarrow p_a/\lambda^2.
 \ee
Therefore it is natural to introduce phase space variables that are invariant under these rescalings. These are 
\be
\label{scaleinv}
A = V_0\,^{1/3} a, \quad \quad p_A = V_0\,^{2/3}p_a;
\ee
the Poisson bracket is then
\be
\{A,p_A \} = 8\pi G,
\ee 
and removes the dependence of the Poisson bracket on the arbitrary coordinate box volume $V_0$. Since this rescaling gives the scale factor the dimension of length, appropriate factors of  $\sqrt{G}$ (or alternatively the Planck length $l_{P}$) may be used to obtain scale invariant and dimensionless phase space variables, and the time parameter. The final variables we use are 
\bea
\label{svars}
\bar{a} &=& \left( \frac{V_{0}}{8\pi}\right)^{1/3} \frac{a}{\sqrt{G}}, \quad
\bar{p}_{\bar a}=\left( \frac{V_{0}}{8\pi}\right)^{2/3} \frac{p_a}{\sqrt{G}} \nn \\
\bar{\phi}&=&\sqrt{8\pi G} \phi, \quad \quad\quad  \bar{p}_{\bar{\phi}}=\left( \frac{V_{0}}{\sqrt{8\pi G}}\right)p_{\phi};
\eea
these lead to the Hamiltonian constraint  
\begin{align}
\label{eq:HamConstSD}
  \mathcal{H}=-\left(\frac{\bar{p}^2_{\bar{a}}} {24\bar{a}}\right)+\frac{\bar{p}^2_{\bar{\phi}}}{\bar{a}^3} \approx0 
\end{align} 
and Friedmann equation
\be 
\label{eq:sffe1}
\left(\frac{\dot{\bar{a}}}{\bar{a}}\right)^2 =\frac{\bar{\rho}}{6}, \quad \bar{\rho}=\frac{\bar{p}_{\bar{\phi}}^{2}}{2\bar{a}^{6}}.   
\ee
In the following we drop the over bars. 

Our goal is to use appropriate functions of the variables (\ref{svars}) for a lattice quantization. To that end we note that the canonical pair
\begin{align}
    q=f(a), \quad p=\frac{p_{a}}{f'(a)},
\end{align}
where $f(a)$ is any once differentiable function, satisfies $\{q,p\}=1$. Since only finite translation and not momentum operators are  definable for quantization on a lattice, it is useful to consider the alternative Poisson bracket 
\be 
\label{gen_pb}
\left\{q, \exp\left(i\lambda p\right)\right\} = i\lambda \exp\left(i\lambda p\right) 
\ee
for quantization; $q$ then determines the type of ``adaptive mesh" configuration lattice with spacing parameter $\lambda$;  each such lattice leads to different physics in the effective quantum theory, as we see below.
%Our goal is to use appropriate functions of the variables (\ref{svars}) for a lattice quantization. Since only finite translation and not momentum operators are  definable for quantization on a lattice, it is useful to consider the alternative Poisson brackets of the form 
%\be 
%\label{gen_pb}
%\left\{f(a), \exp\left[i\mu p_a/f'(a)\right]\right\} = i\mu \exp\left[i\mu p_a/f'(a)\right] 
%\ee
%for quantization, where $f(a)$ is any once differentiable function; this function determines the type of ``adaptive mesh" configuration lattice with spacing parameter $\mu$;  each such lattice leads to different physics in the effective quantum theory, as we see below. Equivalently, it is an alternative choice of canonical variable since $\{f(a),p_a/f'(a)\} = 1$. (Special cases of the function $f(a)$ have been used in the LQG framework applied to cosmology, where the exponentiated variable in (\ref{gen_pb}) is the ``point holonomy" of the Ashtekar-Barbero connection \cite{Agullo:2016tjh}; here $p_a$ arises in symmetry reduction of the ADM momentum). 
For the particular choice 
\be
%\label{eq:sidv1}
q\equiv f(a) = a^n,\quad   p\equiv \frac{p_a}{na^{n-1}}\quad   (n>0)
\label{pq}
\ee
the Hamiltonian constraint takes the form  
\be 
{\mathcal H} =  -\left(\frac{n^{2}}{24}\right) q^{\frac{2n-3}{n}} p^2+\frac{p_{\phi}^2}{2q^{\frac{3}{n}}}\approx 0.
\label{Hpq}
\ee
%where $\rho$ is the scalar field energy density. 
This is the case we consider below for deriving effective equations. 

\section{Lattice Quantization}
\label{QU}
\hspace{\parindent}
 
With the variables $(q,p)$  defined above, the quantization of the Poisson bracket 
 \begin{align}
    \{q,U_{\lambda}(p)\}=i\lambda U_{\lambda}(p), \quad U_{\lambda}(p)=e^{i\lambda p}
\end{align}
is realized on the Hilbert with basis $|q\rangle$ and inner product 
\be
\langle q|q'\rangle = \lim_{T\rightarrow\infty} \frac{1}{2T} \int_{-T}^T d\mu \ e^{-iq\mu}e^{iq'\mu} = \delta_{q,q'}
\ee 
by defining the operators \cite{Halvo,Ashtekar:2002sn}
\be 
\hat{q}|q\rangle= q|q\rangle, \quad \hat{U}_\lambda |q\rangle = |q -\lambda\rangle.
\ee
The resulting commutator algebra is
\be 
[\hat{q},\hat{U}_{\lambda}]=-\lambda\hat{U}_{\lambda}.
\ee
$\hat{U}_{\lambda}\,^{\dagger}=\hat{U}_{-\lambda}$; the parameter $\lambda$ is an arbitrary real number and any particular choice fixes a lattice.

Defining the Hamiltonian constraint (\ref{Hpq}) as an operator requires defining operators for $p^2$ and for powers of $q$, including inverse powers depending on the choice of $n$. The former must be defined using $U_\lambda$; one choice is squaring the momentum operator
\be 
\label{mom-op}
\hat{p}_{\lambda}=\frac{1}{2i\lambda}\left(\hat{U}_{\lambda}-\hat{U}_{-\lambda}\right).
\ee
Negative powers of $q$ may be defined using the type of identity given in  \cite{Thiemann:1996aw}:
\be 
\widehat{\frac{1}{|q|}}=\frac{1}{\lambda^{2}}\left(\hat{U}_{\lambda}\sqrt{|\hat{q}|}\,\hat{U}_{-\lambda}-\hat{U}_{-\lambda}\sqrt{|\hat{q}|}\,\hat{U}_{\lambda}\right)^{2}.
\ee

The standard Schr\"{o}dinger quantization is used for the matter sector since we want to introduce fundamental discreteness for the gravity sector only.

\subsection{Semiclassical states}  

We summarize here an approach used to define effective equations for cosmology that makes use of semiclassical states \cite{HusWin,Tav}. These are Gaussian states of the form $\psi_{\tilde{q},\tilde{p}}(q)\sim e^{-(q-\tilde{q})^2 + i\tilde{p}q}$ where $(\tilde{q},\tilde{p})$ is the point in phase space (defined for the representation given above) on which the state is peaked. With such states suitably defined, the expectation value of the Hamiltonian constraint operator is computed. This gives the effective constraint ${\cal H}_{\rm eff} (\tilde{q},\tilde{p})$, which is then used to derive effective equations by imposing a Poisson bracket on the peaking values $\tilde{q}$ and $\tilde{p}$.

The first step in this procedure is to take the lattice states  $|q_m\rangle = |m\lambda\rangle\equiv |m\rangle_\lambda$, $m\in \mathbb{Z}$. Semiclassical states are defined as in \cite{HusWin}:
\begin{widetext}
\begin{align}
\label{eq:scstate1}
\ket{\tilde{q},\tilde{p};{t,\lambda}}=\frac{1}{C}\sum^{\infty}_{m=-\infty} \exp\left[-\frac{t}{2}\left(m \lambda-\tilde{q}\right)^{2}+im\lambda\tilde{p}\right]\ket{m}_{\lambda},
\end{align}
\end{widetext}
where the normalization constant is 
\begin{align}
C^{2}=\exp\left(-t\tilde{q}^{2}\right)\sum_{m=-\infty}^{\infty}\exp\left[-t\left(m \lambda\right)^{2}+2m\lambda \tilde{q}t\right],
\end{align}
and $t$ is the width of the state and $\left(\tilde{q},\tilde{p}\right)$ are the classical phase space points at which the state is peaked.

Expectation values of the elementary operators in these states are the following:
\bea
\label{expec1}
\lim_{t\rightarrow 0}\big<\hat{q}\big>_{t,\lambda}&=& \tilde{q}\nn\\
\lim_{t\rightarrow 0}\big<\hat{U}_{\lambda}\big>_{t,\lambda}&=&\exp\left(i\lambda \tilde{p} \right)\nn\\
\lim_{t\rightarrow 0}\big<\hat{p}\big>_{t,\lambda}&=&\,\frac{\sin(\lambda \tilde{p})}{\lambda}.
\eea
These are derived using the Poisson resummation formula \cite{HusWin}.  These results are useful for what follows since it bounds the momentum expectation value in the semiclassical state. This is the analog of the ``holonomy correction" in LQC. 

 Lastly the semiclassical state for the scalar field in the $\phi$ representation is taken to be 
 \begin{align}
 \label{eq:scstate2}
\psi_{\tilde{\phi},\tilde{p}_{\tilde{\phi}}}(\phi)=\left(\frac{1}{\pi \bar{t}}\right)^{\frac{1}{4}}e^{-\left(\phi-\tilde{\phi}\right)^{2}/2\bar{t}+i\phi\tilde{p}_{\tilde{\phi}}},
\end{align}
where again  $\bar{t}$ is the width of the Gaussian state and $(\tilde{\phi},\tilde{p}_{\tilde{\phi}})$ are the phase space points on which the state is peaked. This is not defined on a lattice since the idea is to introduce discreteness only for the gravitational variables. 
\section{Effective cosmology}
\label{EC}
%We can define an effective Hamiltonian from the classical expression (\ref{Hpq}) as the expectation value in the semiclassical state of the corresponding constraint operator. 

We define the effective Hamiltonian constraint as the expectation value of the Hamiltonian constraint operator in the semiclassical state. With a symmetric ordering this gives an expression of the type 
\bea
&&\hskip -5mm {\cal H}_{\rm eff}(\tilde{q},\tilde{p}, \tilde{\phi} ,\tilde{p}_\phi;\lambda,t) \nn\\
&&\hskip -5mm =  -\left(\frac{n^{2}}{24}\right) \langle \widehat{q^{\frac{2n-3}{2n}}} \widehat{p}^2_\lambda\widehat{{q}^{\frac{2n-3}{2n}}} \rangle +  \frac{\langle \hat{p}_\phi^2\rangle}{2} \langle \widehat{q^{-\frac{3}{n}}}\rangle.
\eea
This is tedious but straightforward to calculate. A simplification based on the results  (\ref{expec1}) that captures the fact that $\langle \widehat{p}_\lambda\,^2\rangle$ is bounded above is given by the following expression for the effective Hamiltonian constraint in the limit $t\rightarrow 0$:
\bea
\mathcal{H}_{\rm eff}&=& -\left(\frac{n^{2}}{24}\right) q^{\frac{2n-3}{n}}\left[\frac{\sin(\lambda p)}{\lambda}\right]^{2}+\frac{p_\phi^2}{2q^{\frac{3}{n}}}\nn\\
&\approx&0,
\label{HeffF}
\eea
where we have dropped the tilde on the phase space point; the only difference between this expression and the classical one (\ref{Hpq}) is the sine term replacing $p^2$; the arguments used in deriving this are based on the results of \cite{HusWin}, which are similar to those used in the derivation of the LQC effective equation \cite{Tav}.  
 
\subsection{Effective Friedmann equation} 

 Treating the effective Hamiltonian constraint (\ref{HeffF}) as a classical expression we can derive the corresponding canonical equations of motion with lapse $N=1$:  
\begin{align}
    \begin{split}
        \dot{q}=&-\frac{n^{2}}{12\lambda}\,q^{\frac{2n-3}{n}}\sin(\lambda p)\cos(\lambda p),\\
        \dot{p}=&\,\frac{n\left(3-2n\right)}{24}q^{\frac{n-3}{n}}\left[\frac{\sin(\lambda p)}{\lambda}\right]^{2}\\&-\frac{3}{2n}\,p^2_{\phi}\,q^{\frac{-(3+n)}{n}},\\
        \dot{\phi}=&\,p_{\phi}\,q^{-\frac{3}{n}},\quad \quad
        \dot{p}_{\phi}=\,0.
    \end{split}
\end{align}
The square of the Hubble parameter is
\bea
\label{eq:hubsq}
H^{2}&=&\left(\frac{\dot{q}}{nq}\right)^2 \nn\\
&=&\left(\frac{n}{12\,\lambda}\right)^{2}q^{\frac{2(n-3)}{n}}\sin ^{2} \left(\lambda p\right)\nn\\
&&\times\left[1-\sin^{2} \left(\lambda p\right)\right].
\eea
This is simplified by noting that the Hamiltonian constraint (\ref{HeffF}) gives
\begin{align}
\sin ^{2} \left(\lambda p\right)=  
\left(\frac{12\,\lambda}{n}\right)^{2} q^{-\frac{2(n-3)}{n}}\left(\frac{\rho}{6}\right),
\end{align}
where $\rho= p^2_\phi/2q^{6/n}$ is the scalar field energy density. Substituting this expression into (\ref{eq:hubsq}) gives the effective Friedmann equation
\begin{align}
H^{2}=\left(\frac{\rho}{6}\right)\left[1-
q^{-\frac{2(n-3)}{n}}\left(\frac{12\,\lambda}{n}\right)^{2}\left(\frac{\rho}{6}\right)\right].
\label{effF}
\end{align}

In summary so far, we have shown that starting with the phase space variables (\ref{pq}) in the ADM formalism, preforming a lattice (or polymer) quantization of the gravity variables and the standard quantization of the scalar field, and lastly, using the semiclassical states (\ref{eq:scstate1}), yields the effective Friedmann equation (\ref{effF}). Since the l.h.s. is positive, this equation is physically meaningful provided 
\be
\rho\leq 6\left(\frac{n}{12\lambda}\right)^{2}q^{\frac{2(n-3)}{n}},
\ee
and it predicts a bounce when equality holds, i.e. there is a maximum density which depends on the scale factor. 

But this last feature does not make physical sense since a maximum density in quantum gravity should be independent of the scale factor, and should depend only on the Planck density up to an overall dimensionless constant. Hence the effective Friedmann equation is physically meaningful only if $n=3$, i.e. the scale invariant (dimensionless) physical volume $v=V_0\,a^{3}/(8\pi l_{P}\,^{3})$ is used as the configuration variable. This gives the critical density 
\be
\label{rhoC}
\rho _c \equiv \frac{3}{8\lambda^2}
\ee
in Planck units. Recalling that $\lambda$ is dimensionless lattice parameter, the effective Friedmann equation becomes 
\be
H^{2}=\left(\frac{\rho}{6}\right)\left[1-\left(\frac{\rho}{\rho_{c}}\right)\right].
\ee

This result is exactly the same in form as that derived in LQC  using the argument that area is quantized \cite{Agullo:2016tjh} for the elementary plaquette used to define the curvature of the Ashtekar-Barbero connection using holonomies in the Hamiltonian constraint.  Here the steps and argument are different, but lead to the same physical result. 

\subsection{Validity of the effective equation}
\label{EDU}

The effective equations derived above are valid provided quantum fluctuations are small; this is because the regime of large quantum fluctuations should not be describable by effective classical equations composed of a few expectation values. The natural variable for thinking about quantum fluctuations in the $n=3$ case is the physical volume. 

The relevant question is whether volume quantum fluctuations are large at the bounce point, where the critical density is given by (\ref{rhoC}) and the corresponding physical volume (obtained by setting $\rho=\rho_c$) is
\be
v_{\rm bounce} = \frac{p_\phi}{\sqrt{2\rho_c}} = \frac{2\lambda}{\sqrt{3}}\ p_\Phi;
\ee
the scalar field momentum $p_\phi$ is a constant of the motion. For the physically relevant case  $n=3$, the phase space variables are the volume and its conjugate momentum $(v, p_v)$. For these variables the momentum operator as defined in (\ref{mom-op}) and the commutator $\displaystyle \left[\widehat{v},\widehat{U}_{\lambda}\right]=-\lambda\, \widehat{U}_{\lambda} $ lead to the uncertainty relation 
\be 
\label{eq:dvds}
\Delta v\,\Delta \left[\frac{\sin\left(\lambda p_{v}\right)}{\lambda}\right]\geq\frac{1}{2}\big|\big<\cos\left(\lambda p_{v}\right)\big>\big|.
\ee
The effective Hamiltonian constraint (\ref{HeffF}) (with $n=3$) gives
\be
\frac{\sin \lambda p_v}{\lambda } = \frac{2}{\sqrt{3}} \frac{p_\phi}{v};
\ee
substituting this into (\ref{eq:dvds}) gives the quantum fluctuation relation
\be 
\left(\frac{\Delta v}{v}\right)^{2}\geq\frac{\sqrt{3}}{4p_{\phi}}\,\,\big|\cos\left(\lambda p_{v}\right)\big|.
\ee
This shows that minimum uncertainty in the physical volume is controlled by the value of (the constant of motion) $p_\phi$. Therefore  
for minimum uncertainty states, $\Delta v/v \ll 1$ provided 
\be
p_\phi \gg \frac{\sqrt{3}}{4} \big|\cos\left(\lambda p_{v}\right)\big|
\label{Pphi-bound}
\ee
(in Planck units), a result independent of the physical volume $v$. This is the condition of validity of the semiclassical FLRW-scalar equations as derived here, and it is apparently independent of the fiducial volume if the scale invariant scalar momentum is used. However, if we rewrite (\ref{Pphi-bound}) using the non-scale invariant scalar momentum $\tilde{p}_\phi\equiv p_\phi/V_0$, then we find using (\ref{svars}), that 
\be 
\tilde{p}_\phi \gg \frac{\sqrt{3}}{4V_0} \big|\cos\left(\lambda p_{v}\right)\big|;
\ee
this means that it is possible to use a sufficiently large $V_0$ to satisfy the inequality; i.e. $\tilde{p}_\phi$ need not be at the Planck scale. This latter inequality is consistent with the condition derived in \cite{RovEw} 
in the connection-triad variables using a similar analysis.

\section{Summary and discussion}
\label{SaD}
\hspace{\parindent}

Our main goal in this paper was to show that the FLRW model quantized on a lattice in a box using the ADM variables  gives the same result for the effective Friedmann dynamics as that derived in LQC. This was done in several steps: using scale invariant variables for the classical theory that removes box size (the fiducial volume $V_0$),  quantizing using an arbitrary power of the scale factor and its conjugate momentum, using semiclassical states to derive an effective dynamics, and lastly noting that the effective Friedmann equation makes physical sense only for a particular choice of configuration variable: the physical volume of the universe. This means that quantizing on a volume lattice leads to a natural bounce, with the lattice spacing determining the critical density; the box coordinate volume does not appear, and neither does the Immirzi-Barbero parameter of LQG. Despite these differences, our analysis is comparable to that in the connection-triad variables  \cite{Corichi_2008} with the same final result: the volume and its conjugate momentum are the preferred phase space variables if the bounce is to occur at the Planck density.

We also discussed the regime of validity of the effective equations by deriving a condition for volume fluctuations to be small; this requires the scale invariant scalar field momentum to be larger than unity in Planck units, a condition independent of the fiducial volume $V_0$. This may be compared to the analysis in \cite{RovEw}, where the result requires large $V_0$ if the scale non-invariant momentum is used. However our result matches that of \cite{RovEw} for scale invariant variables. We think that the true physical description is provided by variables that are invariant under coordinate rescalings (which are physically unimportant); thus the scale invariant variables are the natural ones to use when quantizing FLRW models in geometrodynamics or connection dynamics.

As noted above, the selection of volume as the relevant configuration variable arises here by imposing the  physical requirement that the bounce occurs at a critical density that is independent of any function of the scale factor, and instead is of the order of the Planck density. In LQC the same choice arises by importing the quantization of area from LQG, and deriving an effective Hamiltonian constraint  by realizing the curvature of the Ashtekar-Barbero connection as a holonomy around a minimal physical area loop; thus, there one could argue that ``quantization of area leads to the bounce," whereas here it comes from requiring that the bounce occur at Planck density.

We conclude by noting that similar lattice quantizations in the ADM variables may be applied with appropriate physical inputs to other cosmological models discussed in LQC, and possibly also to models in spherical symmetry \cite{Husain:2004yz,Husain:2009vx}.

{\bf Acknowledgements} This work was supported by the NSERC of Canada and the FDC award of Centre College. We thank Edward Wilson-Ewing for helpful comments on the manuscript and for pointing out reference \cite{Corichi_2008}. 

\bibliography{eff}

\end{document}